\documentclass[]{spie}  

\usepackage{amsmath,amsfonts,amssymb}
\usepackage{graphicx}
\usepackage{gensymb}
\usepackage{indentfirst}
\usepackage{siunitx}
\usepackage[colorlinks=true, allcolors=blue]{hyperref}

\title{A waveguide spectrometer for high-resolution millimeter-wave imaging}

\author[a]{Annie Tan}
\author[b]{Patrick Ashworth}
\author[b]{Peter S. Barry}
\author[b]{Chris S. Benson}
\author[c]{Karia Dibert}
\author[b]{Harry Gordon-Moys}
\author[d]{Peter Hunyor}
\author[a]{Kirit S. Karkare}
\author[b]{Izaak Morris}

\affil[a]{Boston University, Boston, Massachusetts, USA}
\affil[b]{Cardiff University, Cardiff, Wales, UK}
\affil[c]{California Institute of Technology, Pasadena, California, USA}
\affil[d]{Rutherford Appleton Laboratory Space, UKRI, Didcot, England, UK}

\authorinfo{Corresponding author: Annie Tan (annietan@bu.edu)}

\begin{document} 
\maketitle

\begin{abstract}
We present the design and development of a novel, compact integrated on-chip spectrometer operating at millimeter (mm) and submillimeter (sub-mm) wavelengths. 
The Superconducting Waveguide Integrated Submillimeter Spectrometer (SWISS) promises significant improvements in spectral resolution, optical efficiency, and frequency coverage over current mm-wave spectrometers by using a free-space rectangular waveguide architecture. 
Light is coupled in through the antenna front end, which separates the two polarizations and passes them to the waveguide spectrometer through a contactless metamaterial transition. 
Each waveguide spectrometer consists of a feedline coupled to a series of filters, based on single-pole cavity resonators with varying resonant frequencies. 
To maximize packing density, the filterbanks are oriented out of the focal plane, and an E-plane bend in the output from the cavity filters enables a planar and scalable construction. 
The output of each filter is coupled to a kinetic inductance detector (KID) to enable high multiplexing factors. 
The spectrometer is fabricated using stacks of silicon wafers etched with a deep reactive-ion etch (DRIE) and metallized. 
In these proceedings, we discuss current development on a SWISS prototype, including measurements on a prototype antenna front end contactless transition and fabrication of prototype waveguide spectrometers. 
\end{abstract}


\section{INTRODUCTION}
\label{sec:intro} 
\begin{figure}
    \centering
    \includegraphics[width=.98\linewidth]{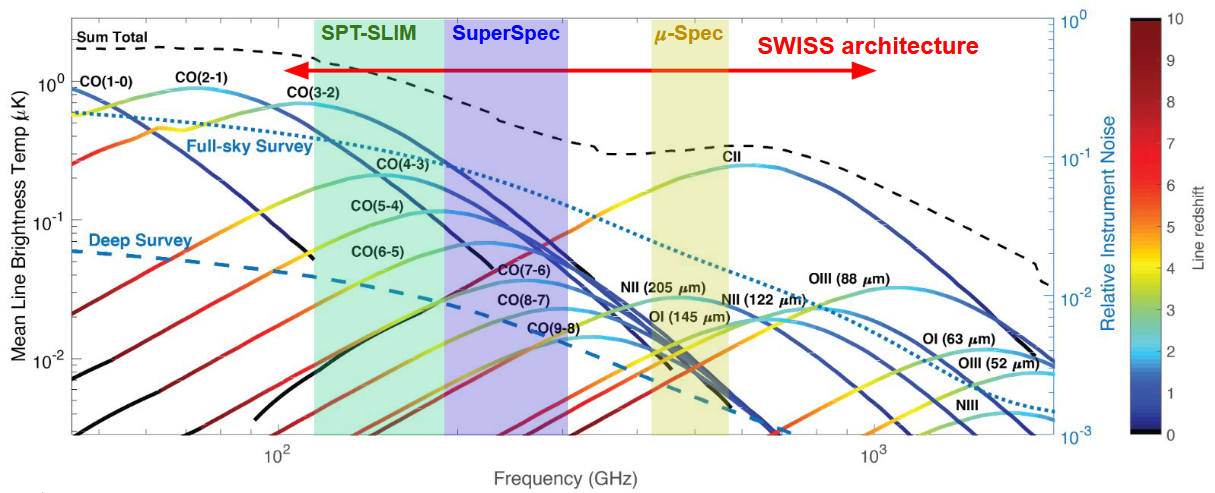}
    \caption{Far-IR lines accessible in the mm/sub-mm regime with source redshift indicated by the color scale. The sensitivity curves are for a 64-spectrometer, R = 300 instrument coupled to an 8.5 K, 3.5 m space telescope with 30\% efficiency observing for 2 years \cite{Delabrouille:2021}. SWISS's architecture would enable an order of magnitude improvement in sensitivity and a wider frequency range than current on-chip spectrometers like SPT-SLIM, SuperSpec, and $\mu$-Spec. Figure adapted from Delabrouille et al.~(2019) \cite{delabrouille2019microwavespectropolarimetrymatterradiation}. }
    \label{fig:snowmassLIM}
\end{figure}

Submillimeter spectroscopy provides access to the high-redshift universe, revealing crucial information about the first stars and galaxies, the history of inflation, the epoch of reionization, and the physics of dark matter and dark energy. 
Dust that absorbs stellar optical and UV light contains atoms and molecules that produce line emission in the millimeter and sub-millimeter bands; of particular interest are the ionized carbon fine-structure line (CII) and the carbon monoxide $J \to J-1$ rotation ladder (CO)\cite{EliVisbal_2010}.
Figure \ref{fig:snowmassLIM} shows line strengths and the wide redshift range available in the submillimeter range.
Line intensity mapping (LIM) is a technique to observe integrated line emission from high-redshift galaxies, detecting 3D large scale structure (LSS) without resolving individual sources \cite{LIMtheory,snowmassLIM}. 
Detecting these lines probes the conditions of the interstellar medium and the physics of star formation and provides redshifts of distant galaxies for LSS mapping in three dimensions.
Studying LSS provides the opportunity to test extensions to $\Lambda$CDM, informing models of dark matter and dark energy.

Several mm-wave spectrometers exist (DESHIMA, SuperSpec, SPT-SLIM, $\mu$-Spec), but they are fundamentally limited in optical efficiency and spectral resolution due to their planar microstrip architecture \cite{Taniguchi2022DESHIMA2,Shirokoff2014SuperSpec,sptslim,muspec}. 
In all current planar filterbank designs, spectral resolution is limited by losses incurred by the dielectric in the microstrip design.
Dissipative losses in the microstrip transporting signal to the detector further limit optical efficiency.
Additionally, current designs are limited in packing density, as most of the available focal plane area is used to house the large planar filterbank. 
Vacuum waveguides potentially solve these problems, and several groups have developed prototype vacuum waveguide spectrometers (W-Spec, Raxdex, NIST)\cite{wspec,raxdex,nist}.
Here we present the Superconducting Waveguide Integrated Submillimeter Spectrometer (SWISS), a scalable prototype vacuum waveguide spectrometer that will significantly improve on spectral resolution, efficiency, and frequency coverage over current spectrometers, designed to be scalable to a densely-packed focal plane.

\section{DESIGN OVERVIEW}
The main difference between SWISS and previous on-chip mm-wave spectrometers is the use of rectangular free-space waveguides. 
Vacuum waveguides do not use a dielectric, which can limit the spectral resolution and optical efficiency in microstrip designs through dissipative losses. 
The waveguides are naturally oriented out of the focal plane, enabling dense packing compared to planar spectrometers.   
SWISS is built around a vacuum waveguide filterbank spectrometer, with all features etched in silicon (Si) wafers, metallized, and then coupled to detectors. 

\begin{figure} [ht]
    \includegraphics[width=.98\linewidth]{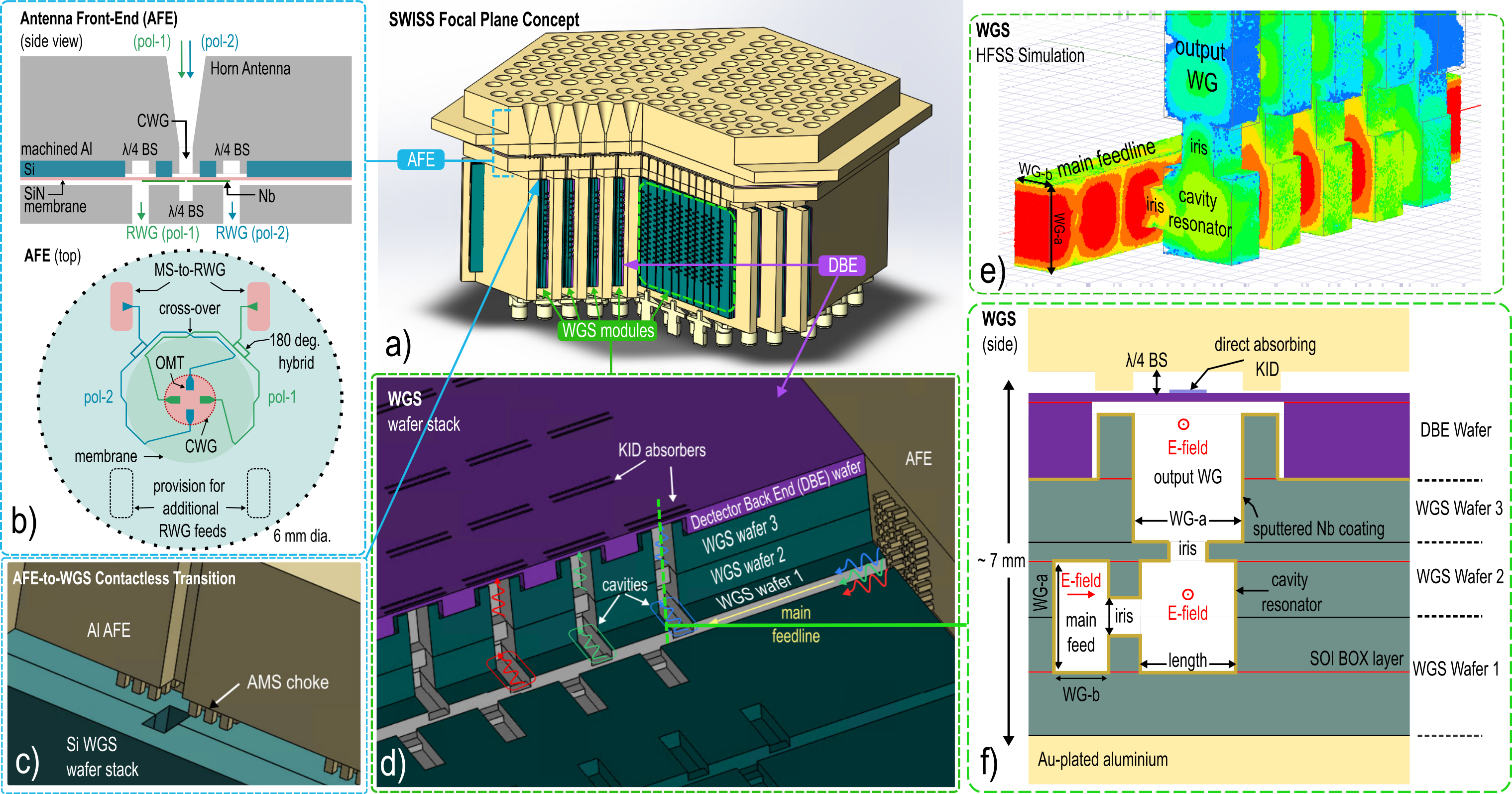}
    \caption{\textbf{(a)} Concept design of a full focal plane array of SWISS modules. \textbf{(b)} Antenna front-end (AFE) schematic, side view and top view. Light enters the horn antenna and the two polarizations are separated by the OMT before being passed to the WGS waveguide. \textbf{(c)} 3D model of contactless AFE to WGS transition in a SWISS module.\textbf{(d)} 3D model of a WGS wafer stack. Broadband signal from the AFE enters the main feedline and encounters a series of tuned cavity resonators, each of which pick off a distinct frequency and sends it to the DBE. \textbf{(e)} HFSS simulation of a 5-resonator spectrometer. \textbf{(f)} Side profile of waveguide spectrometer.}
    \label{fig:main}
\end{figure}

SWISS will integrate all components in one cryogenic module (Figure \ref{fig:main}(a)): antenna front end (AFE), waveguide spectrometer (WGS), and detector back end (DBE).
The AFE couples light in from the sky, separates the orthogonal polarizations, and directs each polarization to an independent spectrometer. 
The WGS contains the waveguide filterbank that separates incident broadband radiation from the AFE into narrow-band channels.
The DBE couples light from each individual WGS channel to a kinetic inductance detector (KID) for readout. 
SWISS will use direct-absorbing KIDs for their sensitivity, multiplexing capability, and scalability for large SWISS arrays. 
The SWISS architecture can be easily scaled for various bands in the mm and sub-mm range, allowing for wider frequency coverage. 

\subsection{Antenna Front End (AFE)}

Light from the sky enters the AFE and is separated into two orthogonal polarizations by a planar orthomode transducer (OMT), with each polarization directed into an independent spectrometer (Figure \ref{fig:main}(b)). 
Each sky pixel consists of a metallic horn that feeds an OMT, which separates the two polarizations in the circular waveguide and routes them to independent microstrip lines that are combined into a \qty{180}{\degree} hybrid coupler. 
Each line is then passed to a probe-coupled microstrip-to-rectangular-waveguide transition that feeds the spectrometer via an output metallic rectangular waveguide on the backside. 

The AFE horn array and output rectangular waveguide will be machined out of aluminum (Al) and coated in gold (Au). 
The output waveguide will be fabricated in an E-plane split-block configuration to allow for precise machining and alignment. 
The OMT will be fabricated on a thin (\qty{2}{\micro\meter}) Si\textsubscript{3}N\textsubscript{4} membrane.

One unique feature of SWISS is the contactless waveguide transition between the AFE output waveguide and the WGS (Figure \ref{fig:main}(c)). 
Since the AFE waveguide will be fabricated with Al and the filterbank waveguide with Si wafers, the two structures will contract at different rates when cooled to sub-Kelvin temperature; a fixed transition between Al and Si would deform and become misaligned. 
We designed a contactless choke to prevent signal leakage between two waveguides even if they are misaligned or present a gap. 

The contactless transition consists of a bed-of-nails structure, which acts as an artificial magnetic conductor (AMC) surface when embedded in a dielectric (vacuum here). 
When a PEC surface is placed on top of the AMC surface with a height less than $\lambda/4$ , no waves can propagate in the gap between the two. 
We designed and tested a prototype AFE contactless transition for the WR6 band (\qtyrange[range-phrase=--,range-units=single]{110}{170}{\giga\hertz}) based on a similar design for WR3 (\qtyrange[range-phrase=--,range-units=single]{220}{325}{\giga\hertz}) by Rahiminejad et al.~(2014)\cite{Rahiminejad_2014}, which is discussed in Section \ref{prototype afe}.

\subsection{Waveguide Spectrometer (WGS)}
The waveguide spectrometer (WGS) is constructed from an array of superconducting waveguide filterbanks (Figure \ref{fig:main}(d)). 
Broadband radiation from the AFE enters the WGS main feedline, encounters a series of cavity resonators, each tuned to pick off a separate narrow frequency band, and is sent to a detector. 
Each cavity is coupled to the main feedline via an E-plane iris on the broad side of the main feedline.

\begin{figure}
    \centering
    \includegraphics[width=0.6\linewidth]{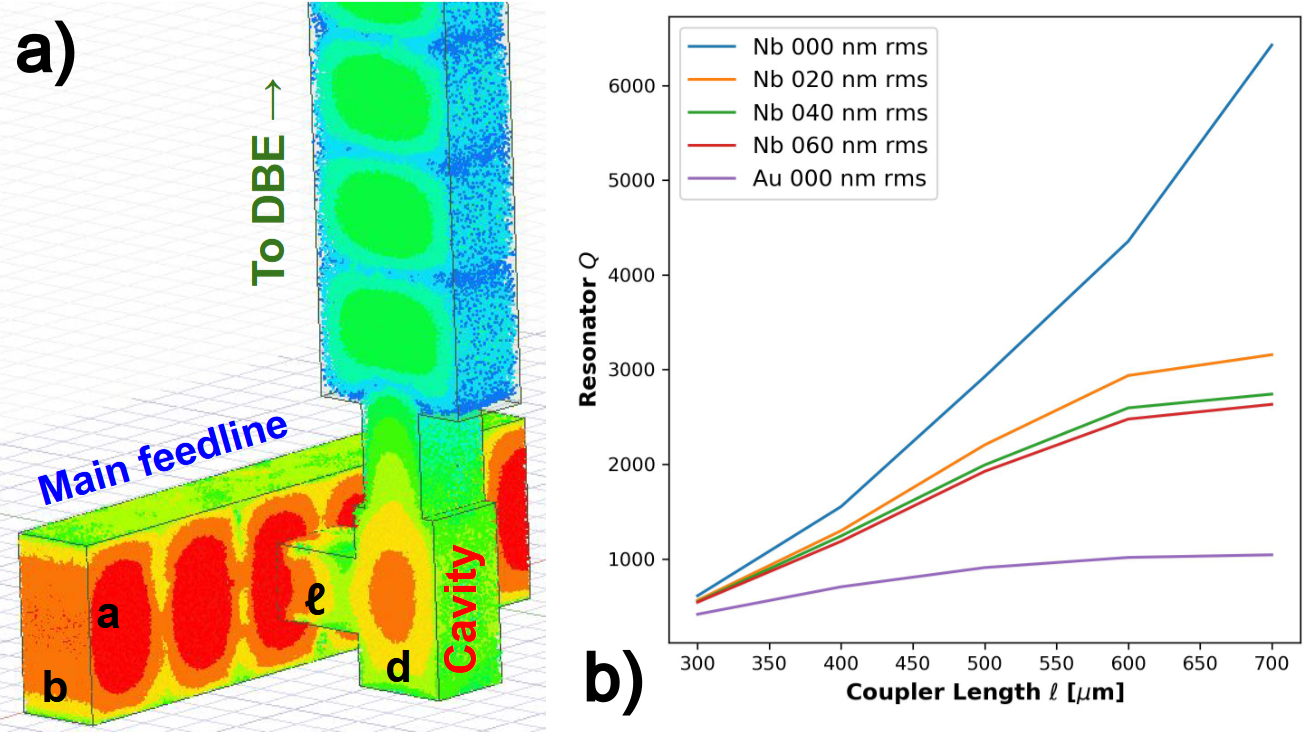}
    \caption{Cavity $Q$-factor optimization for one cavity resonator connected to a feedline and output waveguide. \textbf{(a)} HFSS simulation illustrating a resonator cavity with length $d$ and coupling irises with length $l$. \textbf{(b)} Resonator Q as coupler length $l$ is varied, for superconducting Nb and room-temperature Au. The Nb surface roughness is varied \qtyrange[range-phrase=--,range-units=single]{0}{60}{\nano\meter}. }
    \label{fig:1resoptimization}
\end{figure}

\begin{figure}[h]
    \centering
    \includegraphics[width=.98\linewidth]{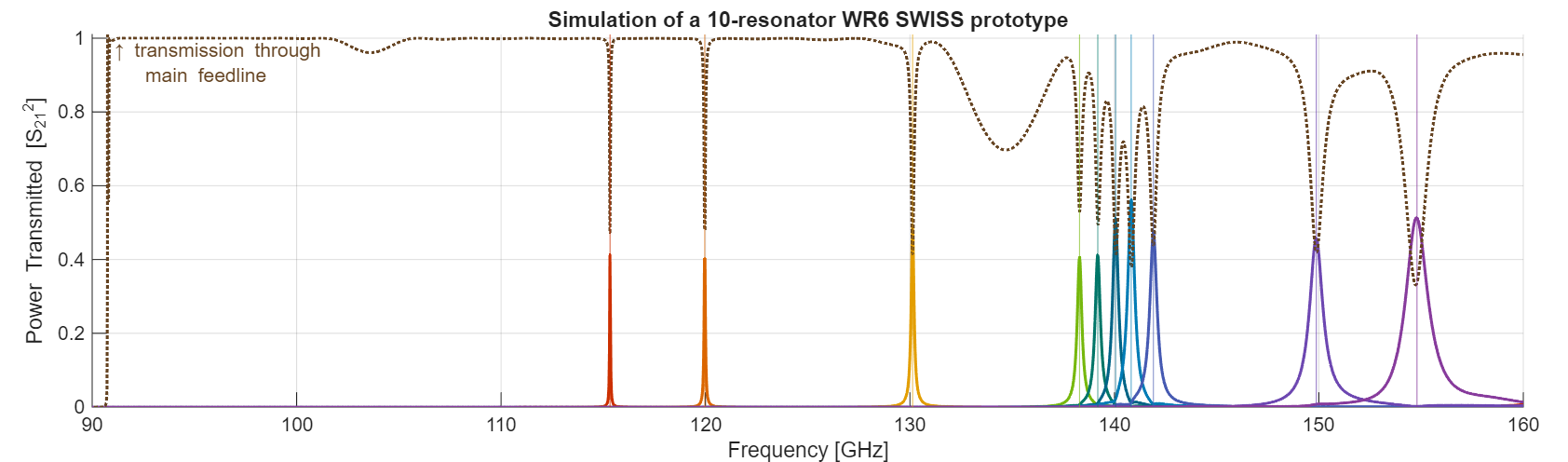}
    \caption{HFSS simulation of a 10-resonator prototype spectrometer, using the same same design as Figure \ref{fig:testing}(a).}
    \label{fig:10chsim}
\end{figure}

The spectral filters are single-pole rectangular cavity resonators. 
To first order, the cavity length $d$ sets the channel frequency $\nu$ with 
\begin{equation}
    \nu_{mnl} = \frac{1}{2\pi\sqrt{\epsilon\mu}}\sqrt{\left(\frac{m\pi}{a}\right)^2 + \left(\frac{n\pi}{b}\right)^2 + \left(\frac{l\pi}{d}\right)^2},
\end{equation}
where $a$ and $b$ are the waveguide dimensions, and $\epsilon$ and $\mu$ are the permittivity and permeability of the material in the cavity (vacuum here)\cite{Pozar2012MicrowaveEngineering}. 
The dominant resonant mode is the TE\textsubscript{101} mode, since $b < a < d$. 
The combination of the coupling iris height and length sets the channel bandwidth, which influences the spectrometer resolution.
The spectral resolution is set by the cavity $Q$-factor, a combination of the iris coupling strength and intrinsic losses: $Q^{-1}_{\rm{res}}=Q^{-1}_{\rm{c}}+Q^{-1}_{\rm{loss}}$.
Figure \ref{fig:1resoptimization}(b) shows $Q$ as the coupling iris length is varied, for both Au and niobium (Nb) coatings.
The WGS rotates the E-field by \qty{90}{\degree} relative to the direction of the main feedline, allowing the output waveguide to be oriented out of plane and providing simple optical coupling to direct-absorbing detectors (Figure \ref{fig:main} (d),(e)). 
Figure \ref{fig:10chsim} shows an HFSS simulation of a SWISS filterbank prototype containing 10 resonators.

The WGS will be assembled with split-block waveguides, split along the E-plane to minimize power leakage and Ohmic losses from defects in the join.
The WGS will be fabricated from a stack of silicon-on-insulator (SOI) wafers that are etched with a deep reactive-ion etch (DRIE), with the buried oxide layer layer of the wafers acting as an etch stop (Figure \ref{fig:main}(f)). 
Since the entire SWISS focal plane module will operate inside a cryostat, the wafers will be metallized with superconducting Nb, which significantly improves $Q$ compared to Au, even with more surface roughness. 
Prototypes for room temperature testing are metallized with Au. 

\subsection{Detector Back End (DBE)}
SWISS detectors will be Al direct-absorbing lumped-element KIDs, which provide high sensitivity, multiplexing capability, and scalability for large SWISS arrays.
Each KID is a superconducting thin-film LC microwave resonator; incoming radiation modifies the inductance of the superconductor, shifting the resonant frequency. 
The inductor in the KID also serves as a tuned mm/sub-mm absorber, which is impedance-matched to the TE\textsubscript{10} waveguide mode. 
SWISS detectors will be similar to those designed for the original SPT-3G+\cite{Dibert_2023}.

\section{PROTOTYPE AFE CONTACTLESS FLANGE} \label{prototype afe}
\begin{figure}[h]
    \centering
    \includegraphics[width=.98\linewidth]{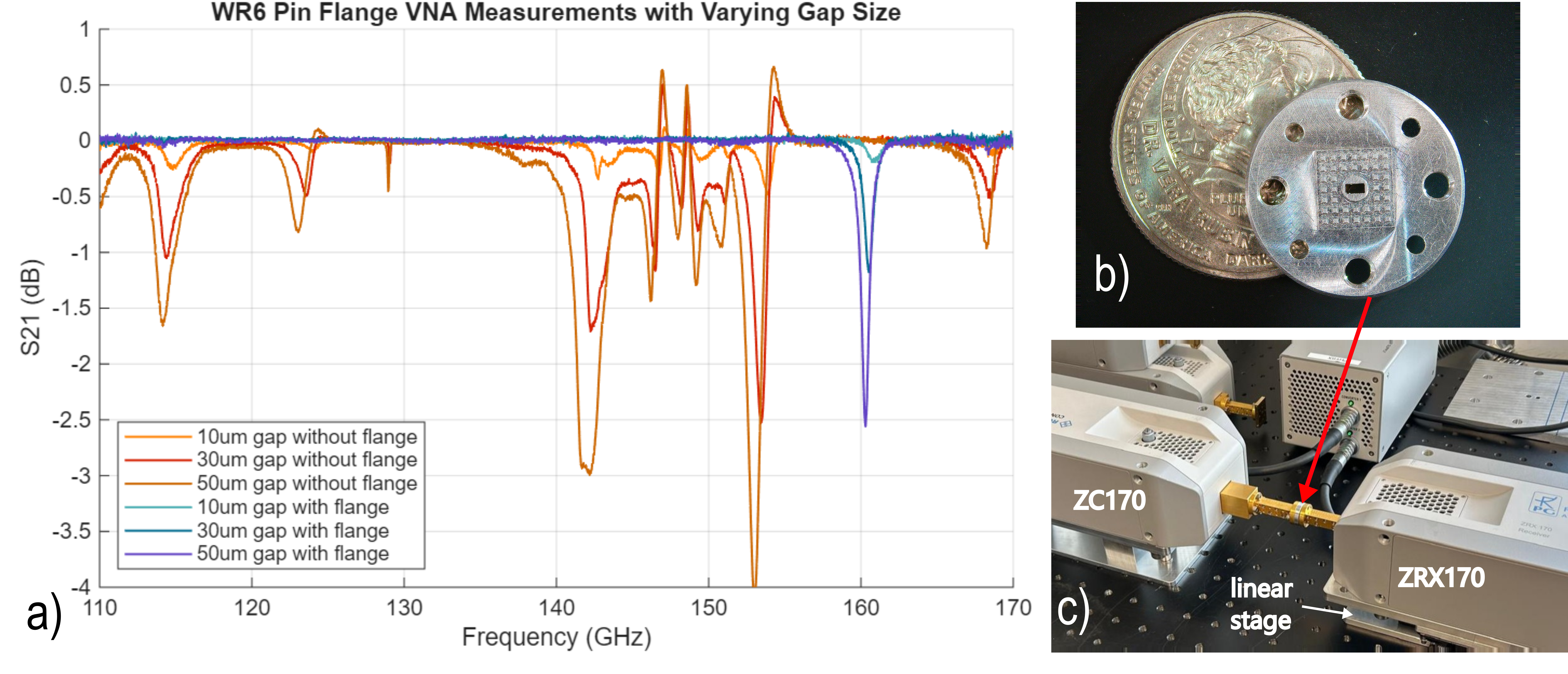}
    \caption{\textbf{(a)} Comparison of transmission ($S_{21}$) results across the WR6 band (\qtyrange[range-phrase=--,range-units=single]{110}{170}{\giga\hertz}), with and without the contactless pin flange while varying the air gap size. Although there is a dip in transmission of \qtyrange[range-phrase=--,range-units=single]{0.1}{2.5}{\decibel} around \qty{160}{\giga\hertz}, we see a similar result in simulation, and it can be moved outside the band by changing pin parameters. \textbf{(b)} Image of the contactless pin flange, US quarter for scale. The rectangular waveguide in the center is surrounded by a circular wall of diameter \qty{2.14}{\milli\meter}, which is surrounded by two rows of seven \qty{0.32}{\milli\meter} wide square pins, spaced \qty{0.59}{\milli\meter} apart. The circular wall and square pins are \qty{0.30}{\milli\meter} tall. \textbf{(c)} The pin flange mounted on the ZC170 frequency converter and aligned with the ZRX170 during VNA measurements. }
    \label{fig:pinflange}
\end{figure}

We manufactured and tested a prototype contactless waveguide flange for the WR6 band to test the designed transition between the AFE and WGS. 
The prototype flange was CNC machined out of aluminum, with a standard WR6 (\qty{1.65}{\milli\meter} $\times$ \qty{0.83}{\milli\meter}) rectangular waveguide opening the center (Figure \ref{fig:pinflange}(b)). 
Two rows of 7 square pins surround the rectangular waveguide opening.
The square pins are \qty{0.32}{\milli\meter} wide and \qty{0.30}{\milli\meter} tall, and the distance between each pin is \qty{0.59}{\milli\meter}. 
A circular wall of diameter \qty{2.14}{\milli\meter} with the same height as the pins surrounds the rectangular waveguide opening. 
The wall acts as an impedance transformer, transforming an open circuit to a short circuit and preventing reflections between the waveguide and the device to be measured. 

This prototype contactless flange was placed between a Rhode \& Schwarz ZC170 Frequency Converter and ZRX170 Receiver, which were both connected to a Rhode \& Schwarz ZVA67 vector network analyzer (VNA) for testing (see Figure \ref{fig:pinflange}(c), VNA not pictured). 
The flange was screwed onto the ZC170 with the pin-side facing the ZRX170.
The ZRX170 was mounted on a linear stage, aligned laterally to the ZC170 using the stage micrometers and aligned vertically using shims. 

We took S\textsubscript{21} measurements using the VNA, with and without the pin flange, while varying the size of the air gap in front of the pins. 
The S\textsubscript{21} curve without the flange and with no gap between the ZC170 and ZRX170 waveguides was used as calibration for all measurements with a gap.
We varied the size of the air gap by moving the ZRX170 with the micrometer in increments of \qty{10}{\micro\meter} and took S\textsubscript{21} measurements.
This set of measurements was repeated with the pin flange screwed onto the ZC170.
A selection of these measurements is shown in Figure \ref{fig:pinflange}(a).
As expected, a larger air gap results in a more lossy transmission profile.
With the pin flange (blue lines), transmission is significantly improved across the band, compared to transmission without the pin flange (red lines).
With the flange, there is a dip in transmission of \qtyrange[range-phrase=--,range-units=single]{0.1}{2.5}{\decibel} around \qty{160}{\giga\hertz}, which matches a dip we found in simulation. 
This dip is an artifact of the pin dimensions and can be adjusted outside of the band.

\section{WGS PROTOTYPES FOR WR3 AND WR6}
Building on the work from Raxdex\cite{raxdex}, we fabricated WR3 and WR6 Si-etched waveguides. 
The fabrication process consists of etching a waveguide trench and resonator pattern into a SOI wafer, metallizing the etched surfaces, and subsequently bonding two identical wafers together to form an enclosed hollow waveguide. 
While the final filterbank devices will be metallized with superconducting Nb for cryogenic operation, Au-metallized waveguides are being fabricated in parallel for room-temperature characterization.

\subsection{Fabrication Details of a WR3 Prototype}
For the initial WR3 prototype, a \qty{3}{\micro\meter} SiO\textsubscript{2} hard mask was deposited onto the SOI wafer. 
The waveguide pattern was then defined using AZ10XT photoresist and transferred onto the hard mask using inductively coupled plasma reactive ion etching (ICP-RIE) with a C\textsubscript{4}F\textsubscript{8}/O\textsubscript{2} plasma chemistry. 
The SiO\textsubscript{2} hard mask was subsequently used for the deep Si etch with a DRIE.

\begin{figure}[h]
    \centering
    \includegraphics[width=\linewidth]{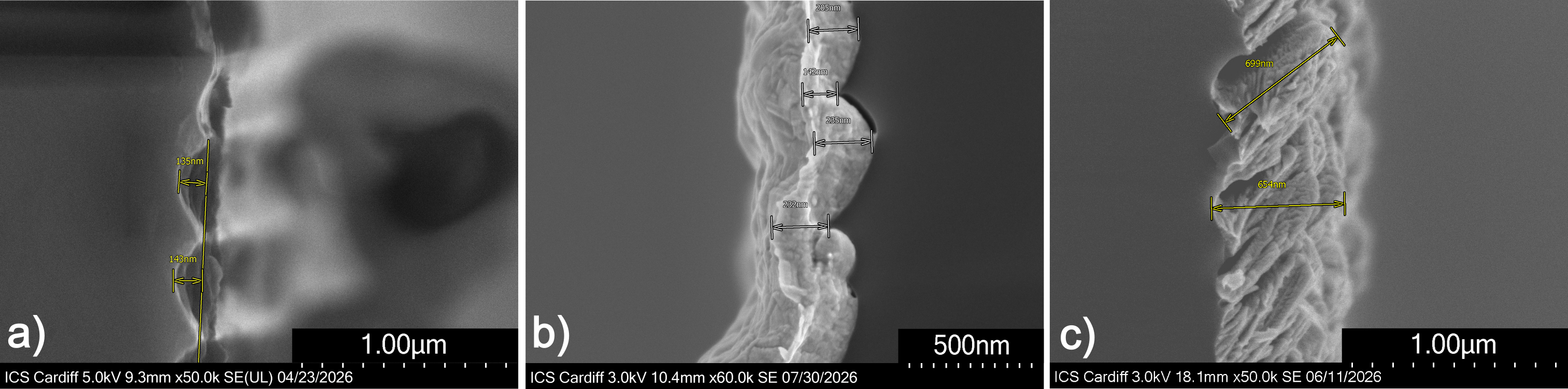}
    \caption{\textbf{(a)} Scallop size of the optimized DRIE recipe, measurements show a surface roughness of $\sim \qty{140}{\nano\meter}$. \textbf{(b)} Au electroplated waveguide wall, film thickness between \qty{142}{\nano\meter} and \qty{235}{\nano\meter}. \textbf{(c)} Nb sputtered waveguide wall, film thickness of $\sim \qty{650}{\nano\meter}$, with angled columnic growth.  }
    \label{fig:WR3SEM}
\end{figure}

DRIE was performed using an SPTS Omega\textsuperscript{\textregistered} DSi-v system (KLA Corp UK Ltd.). 
The Bosch process was optimized to produce \qty{400}{\micro\meter} deep Si cavities with near-vertical sidewalls ($\sim \qty{90}{\degree}$), an etch rate of approximately \qty{9}{\micro\meter\per\minute}, and a Si-to-mask selectivity exceeding 100:1. 
Following process optimization, the sidewall scallop size was reduced to approximately \qty{140}{\nano\meter}, as shown in Figure \ref{fig:WR3SEM}(a). 
We expect that further optimization of the lithography process could reduce the scallop size to below \qty{100}{\nano\meter}.

A key concern identified with this fabrication approach was the scalloping produced by the DRIE process. 
Large scallop amplitudes would compromise metallization of the waveguide walls --- high surface roughness makes continuous coverage within the scallops difficult. 
This becomes increasingly challenging due to the high aspect ratio of the etched trenches. 
To ensure waveguide performance, sufficient material must be deposited to conformally cover the surface variations while also exceeding the required skin depth.

Given the measured sidewall scalloping amplitude of $\sim \qty{140}{\nano\meter}$, the metal layer would need to be sufficiently thicker than this to achieve continuous coverage. 
In additional to full coverage, the minimum thickness would need to exceed the skin depth.
For room-temperature Au in the WR3 band, the minimum thickness required is $\delta_{\rm{Au}}(\qty{220}{\giga\hertz})=\qty{160}{\nano\meter}$.
For superconducting Nb, we find the skin depth is $\delta_{\rm{Nb, SC}}=\qty{129}{\nano\meter}$. 
As a more conservative estimate of the minimum layer of Nb required, we used the room-temperature conductivity of Nb to find $\delta_{\rm{Nb}}(\qty{220}{\giga\hertz}) =\qty{415}{\nano\meter}$. 
Skin depth calculations are detailed in Appendix \ref{skindepth}.

For the room-temperature Au waveguides, the samples were cleaned using a \qty{30}{\minute} acetone ultrasonic bath, followed by a \qty{5}{\minute} isopropyl alcohol (IPA) rinse. 
An O\textsubscript{2} plasma treatment was then performed for \qty{10}{\minute} at \qty{100}{\watt}, followed by a \qty{15}{\minute} piranha solution clean and a \qty{2}{\minute} vacuum hot plate dehydration step.

A seed layer consisting of \qty{20}{\nano\meter} titanium (Ti) and \qty{200}{\nano\meter} Au was deposited using a Moorefield sputtering system. 
The samples were subsequently transferred to RAL Space for gold electroplating. 
Electroplating was performed using a neutral pH (7.0) cyanide-based gold plating solution maintained at \qty{55}{\degreeCelsius}, producing a \qty{99.99}{\percent} purity gold coating on the pre-seeded Si substrates. 
To achieve uniform film coverage and minimize surface roughness, a current density of \qty{0.3}{\ampere\per\square\deci\meter} was applied with continuous agitation of the plating solution.

As shown in Figure \ref{fig:WR3SEM}(b), the electroplating process successfully achieved metallization of the trench sidewalls. 
The cross-sectional analysis indicates an approximately \qty{200}{\nano\meter} thick Au layer at the bottom of the etched trench, demonstrating sufficient coverage of the waveguide cavity surfaces.

The same cleaning procedure was applied to the Nb samples. 
Nb deposition was performed using an in-house sputtering system. 
The process utilized high-speed upstream process gas flow control through mass flow controllers (MFCs), a deposition pressure of \qty{0.1}{\milli\bar}, and pulsed DC sputtering with a pulse duration of \qty{500}{\micro\second}.
As seen in Figure \ref{fig:WR3SEM}(c), we achieved a Nb layer of $>\qty{600}{\nano\meter}$, in excess of the required $\delta_{\rm{Nb}}(\qty{220}{\giga\hertz})$ estimate for room temperature.

From the results shown in Figure \ref{fig:WR3SEM}, the DRIE and metallization processes demonstrate the capability to deposit sufficiently thick metal layers within the etched trenches. 
Both Au and Nb metallization approaches exceed the minimum required deposition thickness and provide coverage of the waveguide surfaces. 
These results demonstrate the feasibility of the proposed fabrication approach and provide confidence for the development of metal-coated vacuum waveguide structures.

A forthcoming paper will provide further details on the fabrication process and additional characterization of the waveguide structures described here. 

\subsection{Fabrication Details of a WR6 Prototype}
In parallel to the WR3 WGS prototype, we fabricated a one-resonator WGS prototype in the WR6 band for room temperature testing. 
The prototype consists of a \qty{40}{\milli\meter} long waveguide connected to a rectangular cavity resonator via a coupling iris.
The cavity resonator is then connected to an output waveguide through another coupling iris.
The main feedline and the output waveguide were both designed to the standard WR6 dimensions of \qtyproduct{1.651 x 0.8255}{\milli\meter}.
To achieve the targeted dimensions with an E-plane split, two \qty{1}{\milli\meter}-thick Si wafers were etched \qty{825}{\micro\meter} and sandwiched together. 
The cavity resonator was lithographically patterned with dimensions of \qtyproduct{1 x 1.573}{\milli\meter} and the coupling irises with dimensions of \qtyproduct{0.2 x 0.519}{\milli\meter}. 

\begin{figure}[h]
    \centering
    \includegraphics[width=.75\linewidth]{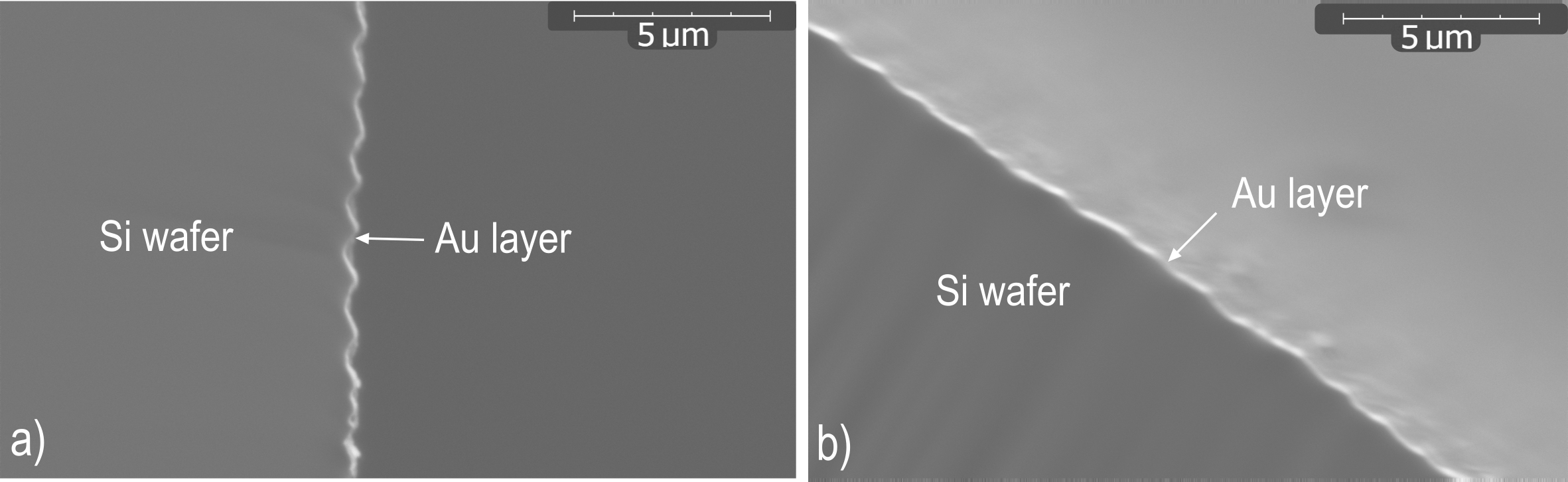}
    \caption{\textbf{(a)} SEM image of a Si wafer etched with the DRIE, followed by deposition of Au using e-beam evaporation. As seen in the image, there is scalloping, with the Au thickness unevenly deposited on the bottom of each scallop. \textbf{(b)} SEM of a Si wafer etched with the DRIE, oxidized in an oxidation furnace, etched with HF, followed by an Au deposition also using e-beam evaporation. Compared to the previous image, the scalloping here is significantly reduced, allowing for a more even Au layer. }
    \label{fig:SEM}
\end{figure}

Figure \ref{fig:testing}(a) shows one of the two Si wafers etched with the prototype pattern; the other wafer contains a mirrored pattern without the coupling irises. 
On the front side, the main feedline, cavity resonator, coupling irises, output waveguide, and dowel holes were patterned with a \qty{11}{\micro\meter} layer of AZ10XT photoresist.
This pattern was etched \qty{825}{\micro\meter} using the three-step Bosch process on the Oxford Estrelas DRIE.
The backside of the wafers were then patterned with the same dowel hole pattern and etched \qty{400}{\micro\meter} to create a through hole.

To achieve the minimum Au thickness of $\delta_{\rm{Au}}(\qty{110}{\giga\hertz})=\qty{226}{\nano\meter}$, the front sides of the wafers were metallized with a \qty{10}{\nano\meter} Ti adhesion layer and a \qty{600}{\nano\meter} layer of Au using electron-beam evaporation on an Angstrom Engineering EVOVAC system. 
Similar to the WR3 prototype, there are also scallops on the vertical sidewalls, with a depth of $\sim \qty{300}{\nano\meter}$ here, preventing complete Au coverage (Figure \ref{fig:SEM}(a)). 
Following the procedure from Jung-Kubiak et al.~(2016) to improve surface roughness\cite{hfetch}, we placed the etched wafer in a thermal oxide furnace to grow \qty{2}{\micro\meter} of SiO\textsubscript{2} and etched this layer away with a hydrofluoric acid (HF) bath, effectively removing the peaks from the scallops. 
Figure \ref{fig:SEM}(b) shows an SEM image of a Si wafer that used this procedure to smooth out the sidewalls, followed with an Au evaporation layer.
Further work is required to ensure sufficient Au coverage $(>\delta_{\rm{Au}})$ across all waveguide surfaces.

\subsection{Room-Temperature Testing Setup for a WR6 Prototype}
\begin{figure}[h]
    \centering
    \includegraphics[width=.7\linewidth]{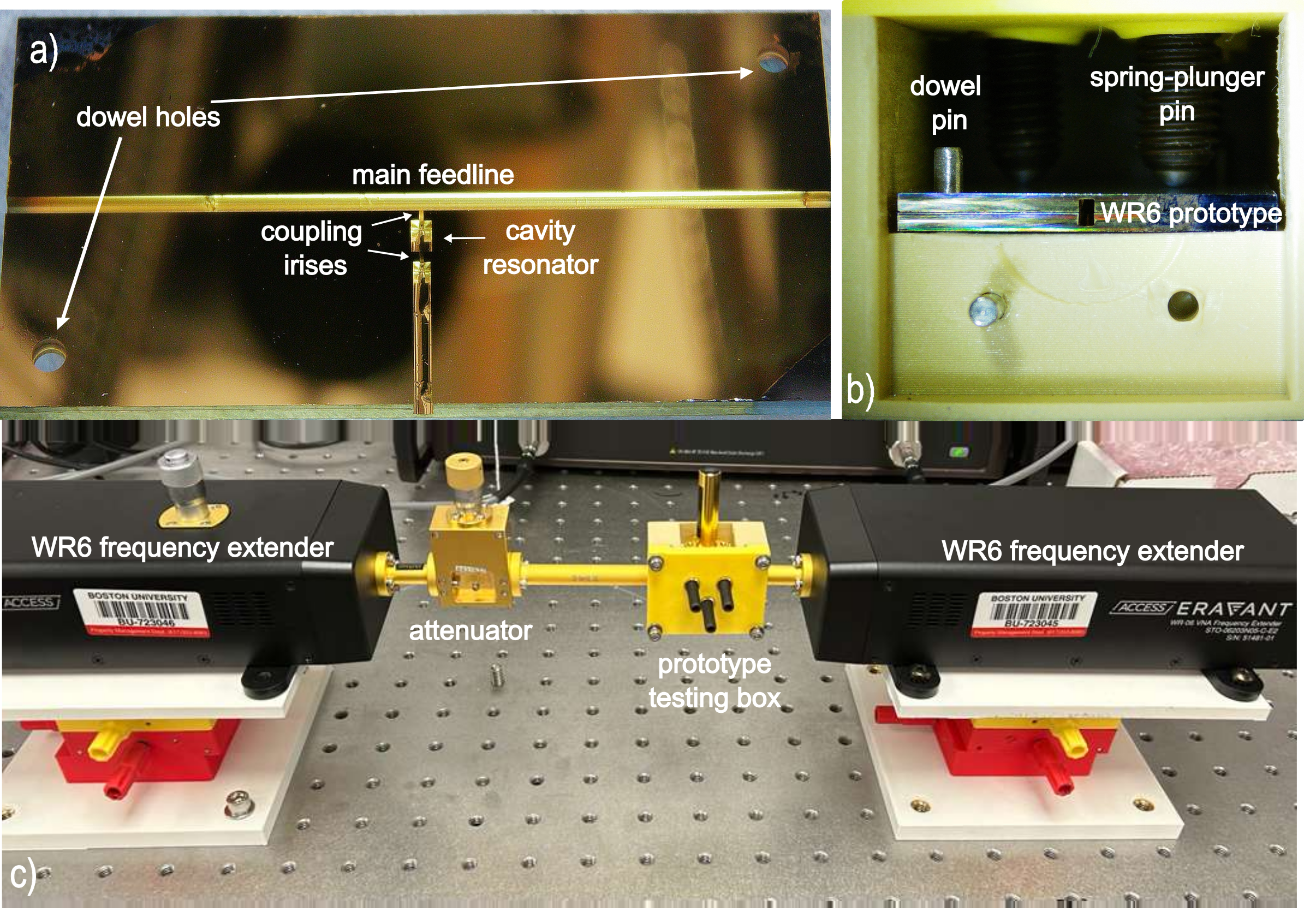}
    \caption{Testing setup for a WR6 prototype. \textbf{(a)} Image of one of the WR6 SWISS prototype wafers, with the waveguide and resonators etched into it and coated in Au. Dowel holes are used for alignment. \textbf{(b)} Side profile of wafer holding box for testing. The two prototype wafers are sandwiched together, aligned with dowel pins and held together with spring-plunger pins. \textbf{(c)} Testing assembly for the WR6 resonator prototype. The wafer holding box is placed in between two WR6 frequency extenders that are connected to a VNA.}
    \label{fig:testing}
\end{figure}

To test the prototype on a VNA, we designed a testing assembly to hold the prototype between two frequency extenders. 
The two wafers are sandwiched together, aligned with dowel pins to within \qty{20}{\micro\meter}, and placed inside a 3D-printed box (Figure \ref{fig:testing}(b)). 
The wafers are held together with three spring-plunger pins. 
The two sides of the box (corresponding to the main feedline) contain dowel pins and holes for alignment with a standard WR6 flange.
The third side of the box (the output waveguide) has similar a similar mating mechanism for a \qty{50}{\ohm} termination load. 
Figure \ref{fig:testing}(c) shows the prototype testing box in the testing assembly. 
Two Eravant WR-06 VNA Frequency Extenders (Tx/Rx Module) are mounted on linear stages and connected to a Copper Mountain Technologies A2202-Fx VNA. 
The frequency extender waveguides are equipped with Eravant Proxi-Flanges\textsuperscript{TM}, which work similarly to the AFE contactless flange we designed and tested in Section \ref{prototype afe}.
One of the frequency extenders (left in Figure \ref{fig:testing}(c)) is connected to a variable attenuator, which is then connected to the prototype testing box via a Proxi-Flange\textsuperscript{TM} straight waveguide. 
The second frequency extender (right in Figure \ref{fig:testing}(c)) is then aligned using the linear stage and connected to the other side of the prototype testing box. 
Our prototype waveguide spectrometer is currently undergoing testing.

\section{CONCLUSION}
We presented the design of SWISS, an integrated vacuum waveguide-based on-chip spectrometer. 
We additionally presented testing results of a prototype AFE-to-WGS contactless waveguide transition and the design and testing setup for a prototype WGS for the WR6 band.
A forthcoming paper will provide further details on the fabrication process and additional characterization of the prototype WGS in the WR3 band. 
Future work involves taking measurements of the room-temperature WR6 WGS prototype and fabricating a superconducting WGS one-resonator prototype that can be coupled to KIDs and tested cryogenically.
In parallel, we plan to develop a 10-resonator filterbank prototype at room temperature, which is simulated in Figure \ref{fig:10chsim}. 

\acknowledgments 
This work was performed using the shared laboratories and instrumentation supported by the Boston University Photonics Center. 
Annie Tan was supported by the Boston University Photonics Center Travel Grant and Graduate Research Abroad Fellowship. 
Support for Karia Dibert was provided by NASA through the NASA Hubble Fellowship grant \#HST-HF2-51573.001.
We thank Nadya Fishchenko and Apollo Albright for helpful discussion on the skin depth calculation of superconducting Nb presented in Appendix \ref{skindepthnb}. 

\appendix
\section{SKIN DEPTH CALCULATIONS}
\label{skindepth}
\subsection{Skin Depth Calculation for Room-Temperature Au}
For Au, we use the standard skin depth formula,
\begin{equation}
    \delta = \sqrt{\frac{2}{\omega\mu\sigma}}\, ,
    \label{normalskindepth}
\end{equation}
where $\omega=2\pi f$ is the frequency, $\mu\approx\mu_0$ is the permeability of Au, and $\sigma_{\rm{Au}}=\qty{4.5e7}{\siemens\per\meter}$ is the conductivity of Au at room temperature.
In the WR3 band (\qtyrange[range-phrase=--,range-units=single]{220}{325}{\giga\hertz}), the minimum skin depth required is $\delta_{\rm{Au}}(\qty{220}{\giga\hertz})=\qty{160}{\nano\meter}$.
In the WR6 band (\qtyrange[range-phrase=--,range-units=single]{110}{170}{\giga\hertz}), we require $\delta_{\rm{Au}}(\qty{110}{\giga\hertz})=\qty{226}{\nano\meter}$.

\subsection{Skin Depth Calculation for Superconducting Nb}
\label{skindepthnb}
For superconducting Nb, we use
\begin{equation}
    \delta = \sqrt{\frac{2}{\omega\mu}}\left(\sqrt{\sigma_1^2+\sigma_2^2}+\sigma_2\right)^{-\frac{1}{2}}\, ,
\end{equation} 
where $\sigma_1$ and $\sigma_2$ are the real and imaginary parts of the complex conductivity, $\sigma=\sigma_1-i\sigma_2$\cite{Dressel_Gruner_2002}. 
Since we will be operating at hundreds of millikelvin, we can use $T\ll T_c=\qty{9.25}{\kelvin}$ to assume $\sigma_2\gg\sigma_1$\cite{klein1994, nuss1991}. 
This allows us to simplify the expression for the skin depth to 
\begin{equation}
    \delta_{\rm{Nb, SC}}=\sqrt{\frac{2}{\omega\mu}}\left(\sqrt{\frac{\sigma_1^2}{\sigma_2^2}+1}+\sigma_2\right)^{-\frac{1}{2}}\approx\frac{1}{\sqrt{\omega\mu\sigma_2}}\, ,
    \label{Nbsimplified}
\end{equation}
where we have used $\sqrt{\frac{\sigma_1^2}{\sigma_2^2}+1}\approx1$.
At $\hbar\omega\ll2\Delta$, where $\Delta$ is the superconducting gap, 
\begin{equation}
    \frac{\sigma_2}{\sigma_n} = \frac{\pi\Delta}{\hbar\omega}\tanh{\frac{\Delta}{2kT}}\, ,
\end{equation}
where $\sigma_n=\qty{6.7e6}{\siemens\per\meter}$ is the normal state conductivity of Nb\cite{tinkham2004}. In the $T\ll T_c$ limit, this becomes
\begin{equation}
\frac{\sigma_2}{\sigma_n} = \frac{\pi\Delta}{\hbar\omega}\, .
\label{sigma2}
\end{equation}
Plugging $\sigma_2$ from Equation \ref{sigma2} into Equation \ref{Nbsimplified}, we find
\begin{equation}
    \delta_{\rm{Nb, SC}}\approx\sqrt{\frac{\hbar}{\pi\mu\Delta\sigma_n}}\, .
\end{equation}
Since $T\ll T_c$, we use the zero temperature superconducting gap value, $\Delta(0)=\qty{1.5}{\milli\electronvolt}$, to find a skin depth of $\delta_{\rm{Nb, SC}}\approx\qty{129}{\nano\meter}$\cite{Arnold1980ProximityII, Townsend1962EnergyGaps,Sherrill1961SuperconductingTunneling}.

For room-temperature Nb, we can use Equation \ref{normalskindepth} with the normal state conductivity $\sigma_n$ to find $\delta_{\rm{Nb}}(\qty{220}{\giga\hertz})=\qty{415}{\nano\meter}$ for WR3 and $\delta_{\rm{Nb}}(\qty{110}{\giga\hertz})=\qty{586}{\nano\meter}$ for WR6.
\bibliography{report} 
\bibliographystyle{spiebib} 

\end{document}